\documentclass{webofc}
\usepackage{lineno}
\usepackage[varg]{txfonts}   

\begin{document}

\title{Production of light nuclei in small collision systems with ALICE}

\author{\firstname{Chiara} \lastname{Pinto}\inst{1,2}\fnsep\thanks{\email{chiara.pinto@cern.ch}} 
       on behalf of the ALICE Collaboration
}

\institute{Dipartimento di Fisica e Astronomia "E. Majorana", Universit\`{a} degli Studi di Catania,  Via S. Sofia 64, Catania, Italy
\and
           INFN Sezione di Catania, Via S. Sofia 64, Catania, Italy
          }

\abstract{%
The energy densities reached in high-energy hadronic collisions at the LHC allow significant production of light (anti)nuclei. Their production yields have been measured as a function of $p_{\rm T}$ and charged-particle multiplicity in different collision systems and at different center-of-mass energies by \mbox{ALICE}. One of the most interesting results obtained from such a large variety of experimental data is that the dominant production mechanism of light (anti)nuclei seems to depend solely on the event charged-particle multiplicity. Evidence for this comes from the continuous evolution of the deuteron-to-proton and $^3$He-to-proton ratios with the event multiplicity across different collision systems and energies. The characterization of the light nuclei production mechanism is complemented by measurements of their production yields in jets and in the underlying event. 
In this paper, recent results on light nuclei production in small collision systems are shown and discussed in the context of the statistical hadronization and coalescence models. In addition, recent results on the deuteron production in jets and new preliminary results on its production in the underlying event measured in pp collisions at $\sqrt{s} = $ 13 TeV are discussed.
}
\maketitle
\section{Physics motivation}

The production mechanism of light (anti)nuclei is under intense debate in the heavy-ion physics community. 
Currently, the measured production yields can be described by two classes of models: the thermal-statistical models and the coalescence ones. 
In the Statistical Hadronization Model (SHM) \cite{SHM}, hadrons are produced by a thermally and chemically equilibrated source and their abundances are fixed at the chemical freeze-out. This model provides a good description of the measured hadron yields in central A--A collisions \cite{SHM_2}.
However, the mechanism of hadron production and the propagation of loosely-bound states through the hadron gas phase are not addressed by this model. 
On the other hand, the production of light (anti)nuclei can be modelled via the coalescence of protons and neutrons that are close by in phase space at the kinetic freeze-out and match the spin, thus forming a nucleus \cite{coalescence}. The key parameter of the coalescence models is the coalescence parameter $B_{\rm A}$, which is related to the production probability of the nucleus via this process and can be calculated from the overlap of the nucleus wave function and the phase space distribution of the constituents via the Wigner formalism \cite{coalescence_theory}.

\section{Ratio of nucleus and proton integrated yields}
Light (anti)nuclei are identified using the detectors of the central barrel, which cover the pseudorapidity window $|\eta|<$ 0.9. Specifically, the Time Projection Chamber (TPC) specific energy loss (d$E$/d$x$) allows the separation from other particles of deuterons in the low transverse momentum region and of nuclei with Z=2 in the full $p_{\mathrm T}$. The use of the Time of Flight (TOF) information complements the light nuclei identification in the high $p_{\mathrm T}$ region.
In order to extract the light (anti)nucleus integrated yields, $p_{\mathrm T}$ spectra are extrapolated to the unmeasured regions by means of a fit with a Lévy-Tsallis function \cite{LeviTsallis}.

\begin{figure}[!hbt]
\centering
\begin{minipage}[b]{0.49\textwidth}
	\includegraphics[width=\textwidth]{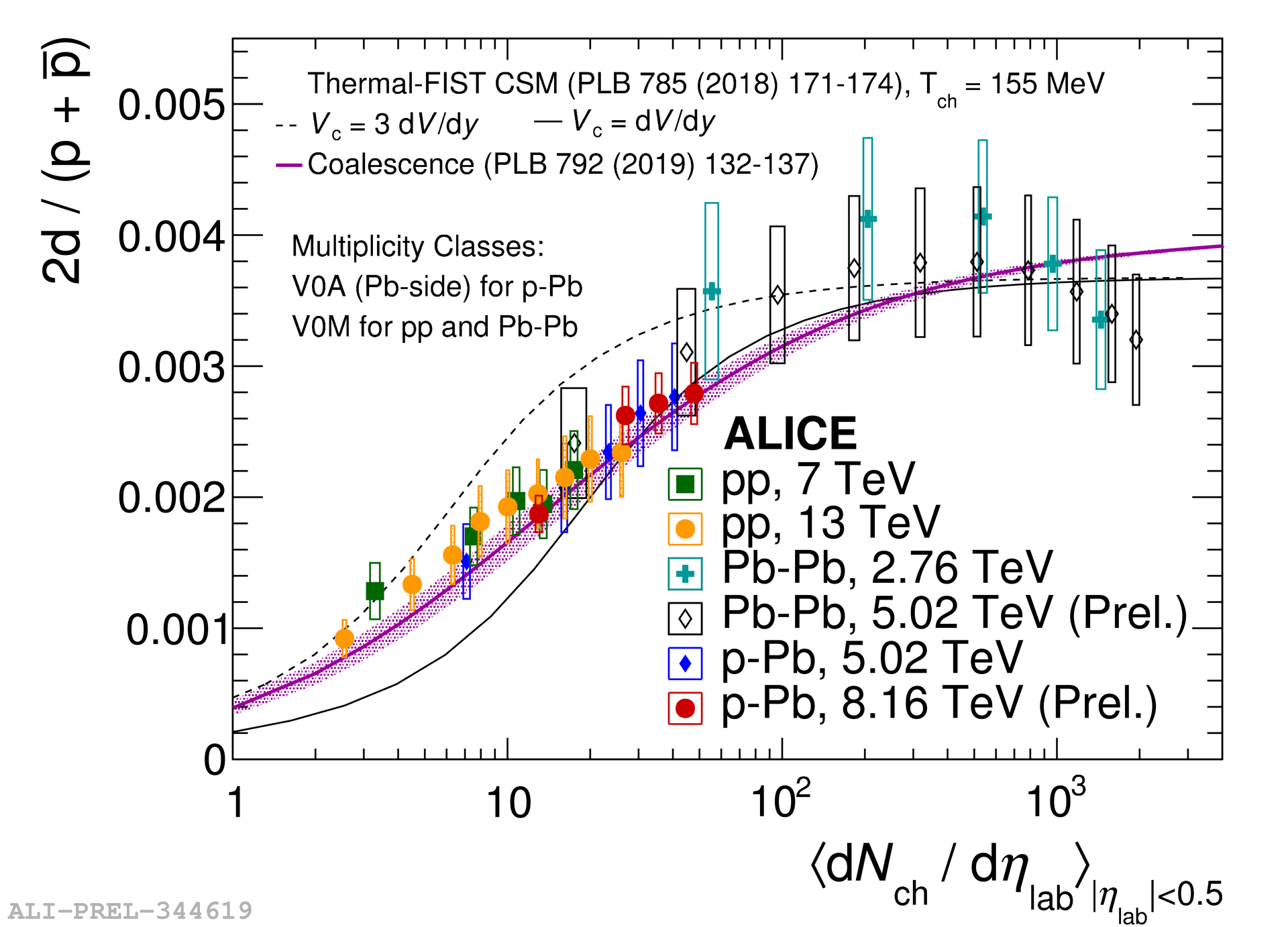}
\end{minipage}
  \hfill
\begin{minipage}[b]{0.49\textwidth}
    \includegraphics[width=\textwidth]{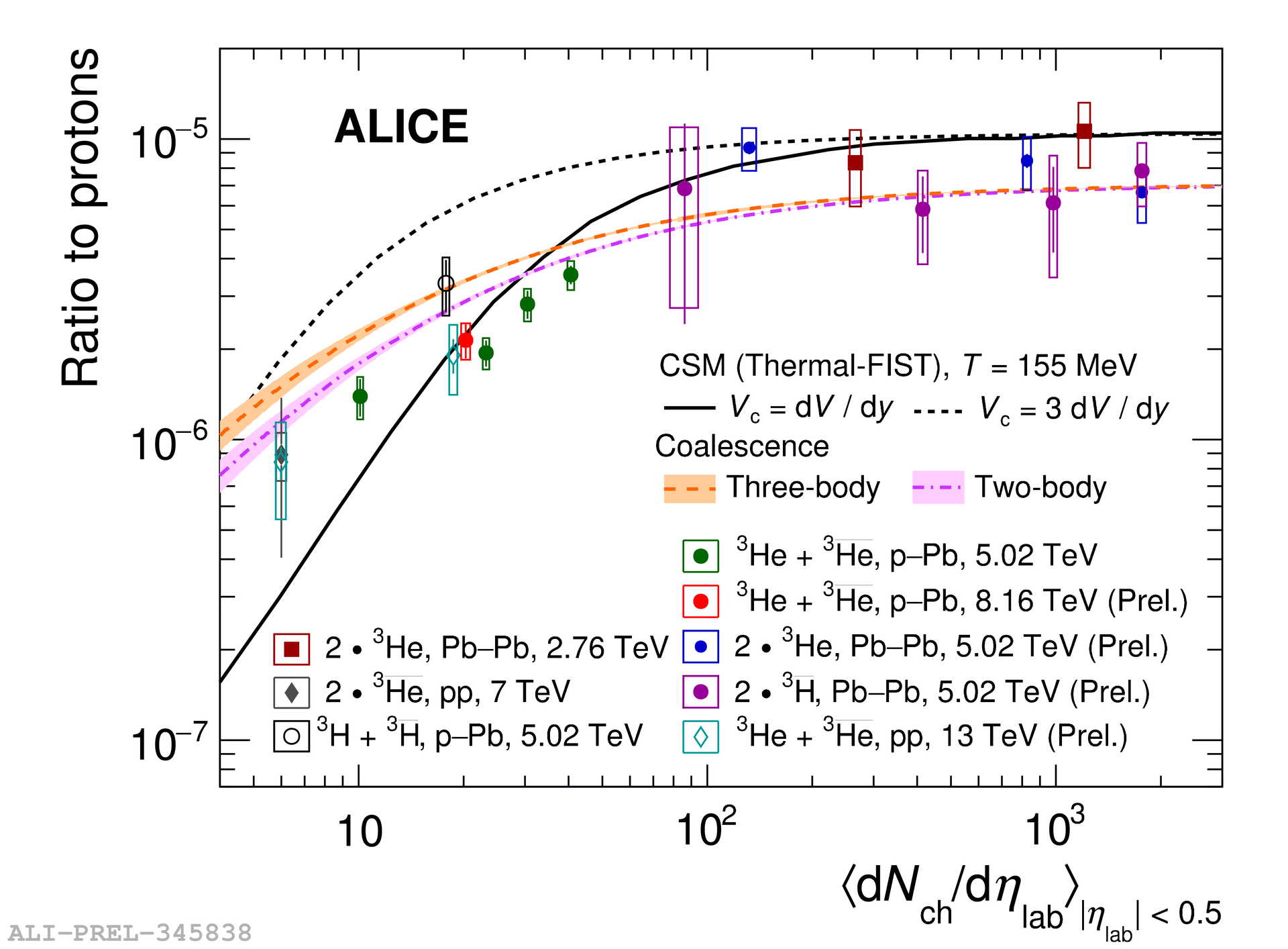}
\end{minipage}   
\caption{Deuteron (left panel), (anti)$^3$He and (anti)$^3$H (right panel) over proton integrated yield ratios as a function of the mean charged-particle multiplicity density for different collision systems and energies. Lines show the statistical uncertainties whereas boxes represent the systematic ones. The results are compared to the expectations of SHM and coalescence models described in the legends. }
\label{fig:NucleiOverP}
\end{figure}

\noindent The ratio between the measured yields of nuclei and that of protons is sensitive to the particle production mechanism. In Fig. \ref{fig:NucleiOverP} the ratio between deuteron (left panel), $^3$H and $^3$He (right panel) measured yields and proton yields as a function of the mean charged-particle multiplicity density ($\langle$d$N_{\mathrm{ch}}/$d$\eta_{\mathrm{lab}}\rangle$) measured in pp, p--Pb and Pb--Pb collisions \cite{pp7TeV, pPb5TeV, ppPbPb, ppSeveralEnergies,pp13TeV,He3pPb5TeV} is compared to the expectations of the models. These nuclei-to-proton yield ratios increase smoothly with the multiplicity, reaching constant values in Pb--Pb collisions. The two ratios show a similar trend with $\langle$d$N_{\mathrm{ch}}/$d$\eta_{\mathrm{lab}}\rangle$, suggesting that the production mechanism depends solely on the event charged-particle multiplicity density. The observed evolution of the d/p ratio is well described by the coalescence approach across all multiplicities. 
For high $\langle$d$N_{\mathrm{ch}}/$d$\eta_{\mathrm{lab}}\rangle$, the coalescence calculations and the canonical statistical model (CSM) expectations are close and both describe the behavior observed in data. On the other hand, the ratio to protons for nuclei with $A$ = 3 is described only qualitatively by both models.
Therefore, in the current state it is not possible to discern between the production mechanisms of light nuclei. In order to further investigate the phenomena underlying the production of light nuclei, it is interesting to study the small system properties by means of new observables, such as the underlying event (UE) activity \cite{UE}.

\section{Underlying Event activity}
ALICE recently released a new analysis performed on pp data
collected at a center-of-mass energy $\sqrt{s}$ = 13 TeV, aiming at understanding the possible role played by the underlying event activity in the light nuclei production. The transverse plane is divided into three azimuthal regions, identified by the relative position with respect to the highest-$p_{\mathrm T}$ hadron (trigger particle). 
The three equal-size regions, $\pi/3$ wide, are: the one around the trigger particle (towards), the one back-to-back to it (away), and the one transverse to both of them, where the UE contribution is dominant. The activity in the UE can be quantified using the self-normalised charged-particle  multiplicity in the transverse region $R_{\mathrm T} = \frac{N_{ch, \mathrm T}}{\langle N_{ch, \mathrm T} \rangle}$ \cite{UE}. Only events with at least one leading charged particle with $p_{\mathrm T} \geq$ 5 GeV/$c$ are considered, since the particle density in the transverse region reaches a plateau at LHC energies. 
Small $R_{\rm T}$ values correspond to an event topology dominated by the hard scattering, while large $R_{\rm T}$ values correspond to events characterized by large UE activity. 

\section{Coalescence parameters}
The coalescence parameter $B_{A}$ is given by the ratio between the invariant yield of the nucleus with mass number $A$ and that of protons, defined as:

\begin{equation}
B_{A} = { \biggl( \dfrac{1}{2 \pi p^{A}_{\mathrm T}} \dfrac{ \mathrm{d}^2N_A}{\mathrm{d}y\mathrm{d} p_{\mathrm T}^{A}}  \biggr)}  \bigg/{  \biggl( \dfrac{1}{2 \pi p^{\rm p}_{\mathrm T}} \dfrac{\mathrm{d}^2N_{\mathrm p} }{\mathrm{d}y\mathrm{d}p_{\mathrm T}^{\mathrm p}} \biggr)^A}  ,
\label{eq:BA}
\end{equation}

\noindent where the labels $A$ and p indicate the nucleus and the proton, respectively, and \mbox{$p_{\mathrm T}^{\rm p}$ = $p_{\mathrm T}^{A}$/$A$}. 

\begin{figure}[!hbt]
\centering
\begin{minipage}[b]{0.48\textwidth}
	\includegraphics[width=\textwidth]{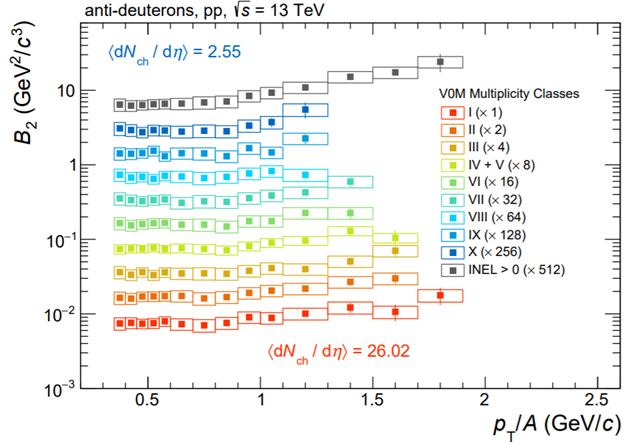}
\end{minipage}
  \hfill
\begin{minipage}[b]{0.49\textwidth}
    \includegraphics[width=\textwidth]{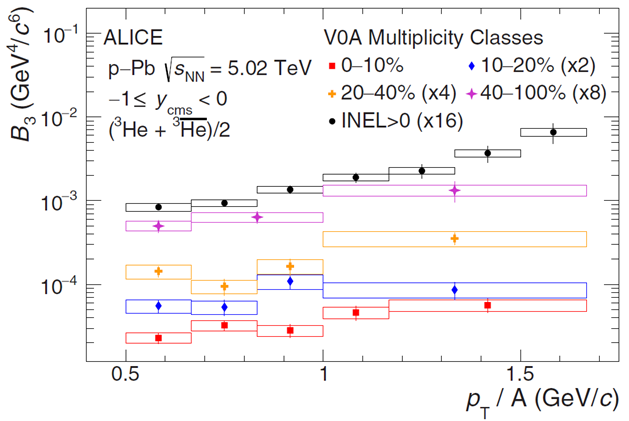}
\end{minipage}   
\caption{$B_2$ as a function of $p_{\mathrm T}$/$A$ in pp collisions at $\sqrt{s}$ = 13 TeV (left panel) \cite{pp13TeV} and $B_3$ in p--Pb collisions at $\sqrt{s_{\mathrm{NN}}}$ = 5.02 TeV (right panel) \cite{He3pPb5TeV}.}
\label{fig:CoalescenceParameters}
\end{figure}

The coalescence parameter $B_A$ has been measured as a function of $p_{\mathrm T}/A$ in different collision systems. In Fig. \ref{fig:CoalescenceParameters}, $B_2$ in pp collisions at $\sqrt{s}$ = 13 TeV and $B_3$ in p--Pb collisions at \mbox{$\sqrt{s_{\mathrm{NN}}}$ = 5.02 TeV} are shown for several multiplicity classes and in the minimum bias (MB) case.
$B_2$ in pp collision is flat with $p_{\mathrm T}/A$ in all multiplicity classes, in agreement with the predictions of a simple coalescence model, which takes into account only momentum space distributions and not space-time correlations \cite{B2flatinpT}. The rising trend of $B_A$ observed in the MB case can be explained as a consequence of the hardening of the proton spectra with increasing multiplicity in addition to hard scattering effects at high $p_{\rm T}$. $B_3$ in \mbox{p--Pb} collisions is slightly increasing with $p_{\mathrm T}$ and it is known from previous measurements that the $B_A$ parameters in Pb--Pb collisions also increase as a function of transverse momentum \cite{ppPbPb}. The behavior of $B_3$ with $p_{\mathrm T}/A$ in p--Pb collisions cannot be explained by simple coalescence hypotheses, as investigated in Ref. \cite{He3pPb5TeV}. Therefore, the observed results, in addition to what is known from earlier measurements, suggest that more sophisticated coalescence models have to take into account the volume dependence in order to explain the data. 

\begin{figure}[!hbt]
\centering
\begin{minipage}[b]{0.49\textwidth}
	\includegraphics[width=\textwidth]{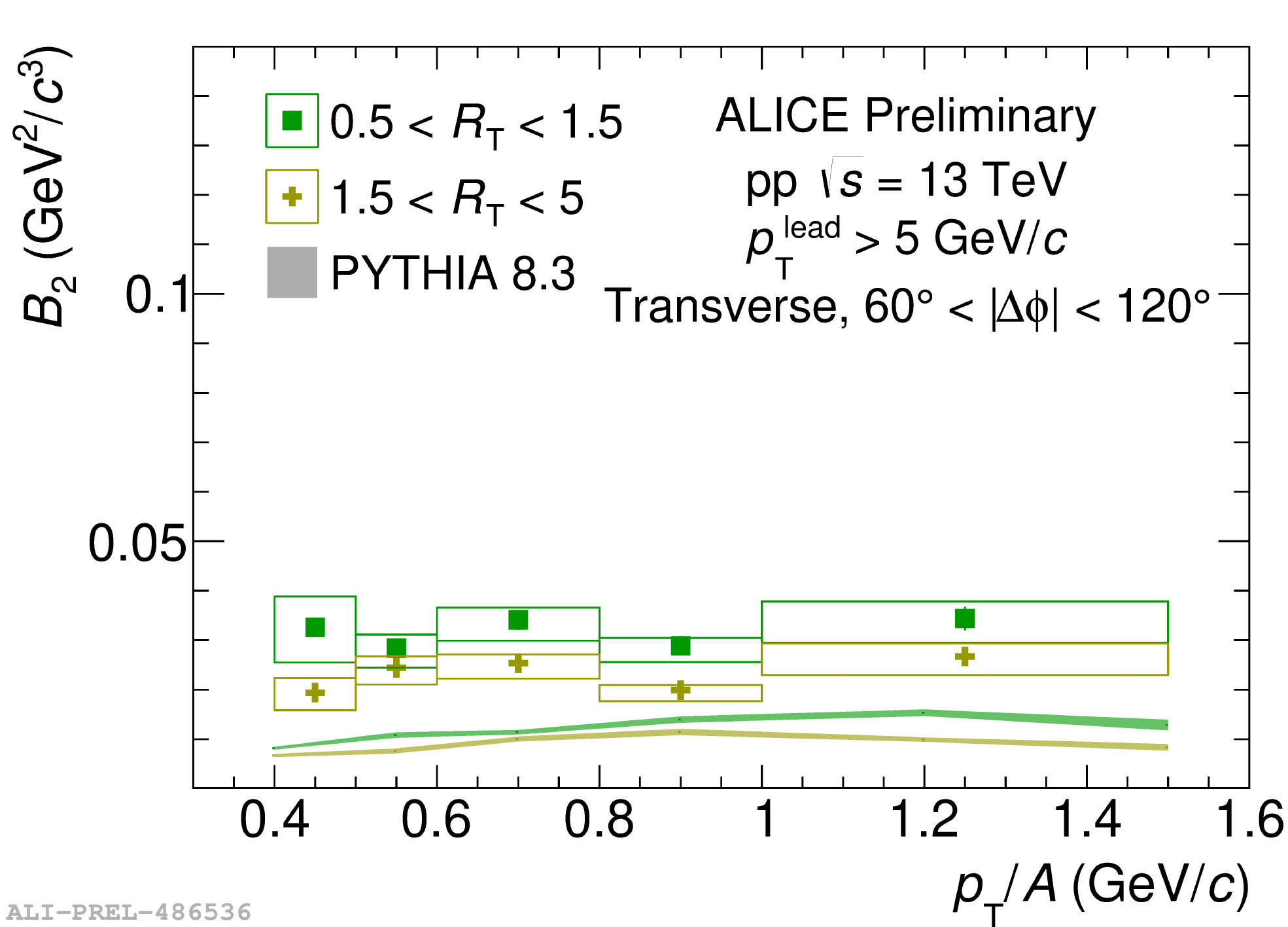}
\end{minipage}
  \hfill
\begin{minipage}[b]{0.49\textwidth}
    \includegraphics[width=\textwidth]{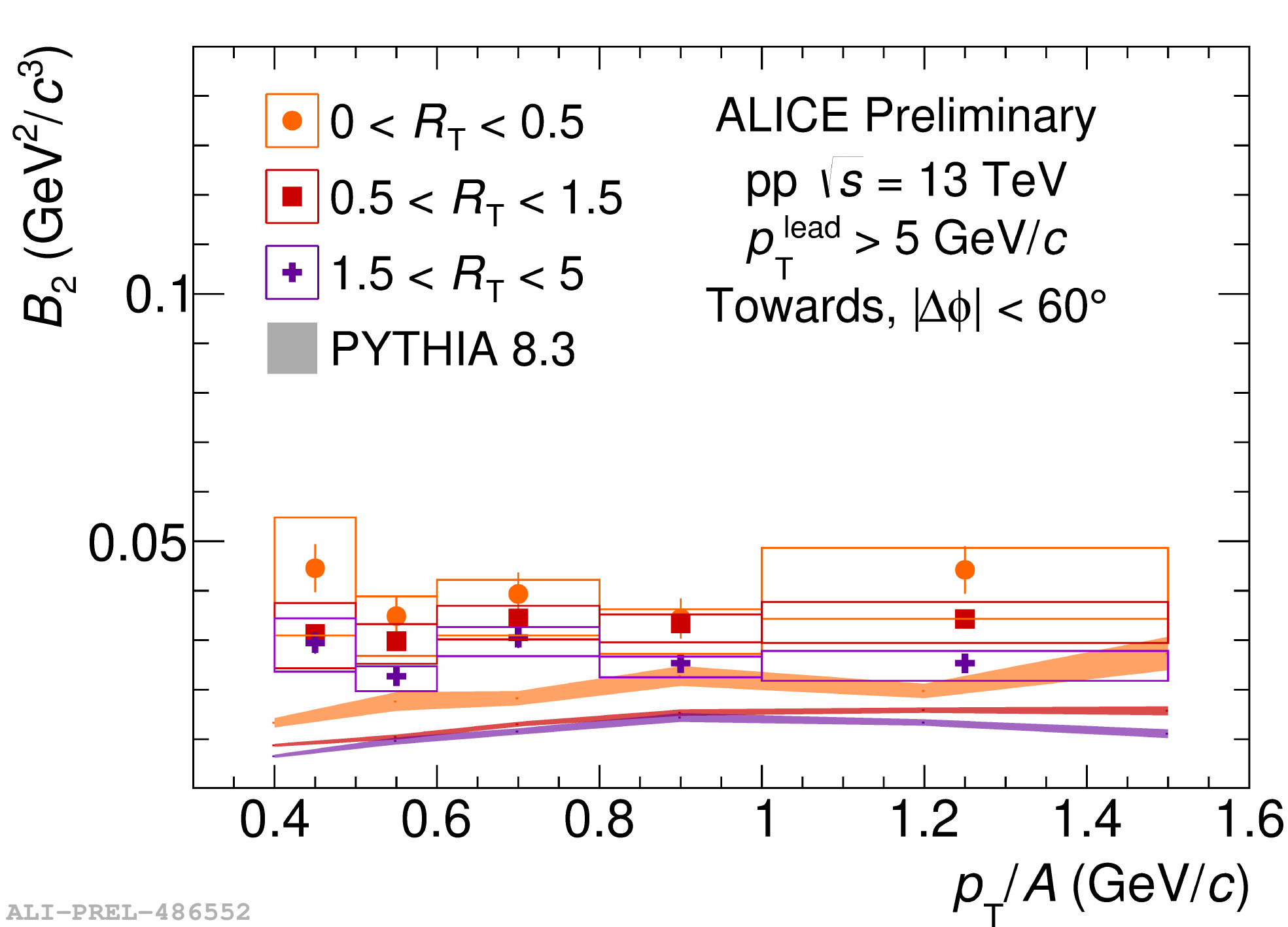}
\end{minipage}   
\caption{$B_2$ as a function of $p_{\mathrm T}$/$A$ in pp collisions at $\sqrt{s}$ = 13 TeV, for several $R_{\mathrm T}$ classes, in the transverse region on the left and in the towards region on the right. }
\label{fig:CoalescenceParametersRT}
\end{figure}

In Fig. \ref{fig:CoalescenceParametersRT}, $B_2$ measured in pp collisions at $\sqrt{s}$ = 13 TeV in the transverse and towards regions, for several $R_{\mathrm T}$ classes, are compared with the expectations of Pythia 8.3 simulations. The $B_2$ parameters as a function of $R_{\mathrm T}$ are flat with $p_{\mathrm T}/A$ in all the azimuthal regions, in agreement with the predictions of a simple coalescence picture.
However, similar values of $B_2$ are found in the transverse and towards regions, against any naive expectations that would expect a larger $B_2$ in the towards region, where nucleons are closer due to the presence of a jet, with respect to what happens in the transverse region. Pythia 8.3 simulations include deuteron production via coalescence considering the reaction cross sections \cite{pythia}. The $p_{\mathrm T}$-dependence and the $R_{\mathrm T}$ ordering are well reproduced by Pythia simulations. 
However, Pythia fails in reproducing the magnitude of the coalescence parameter, therefore further tuning of the Monte Carlo parameters are needed to reproduce the data. 

The obtained results suggest that deuterons are mostly produced in the underlying event. The $B_2$ values are close in all azimuthal regions indicating that the predominant contribution in the deuteron production is due to the underlying event particle production. 
In support of this argument, the recent results of deuteron production in jets \cite{dinjets} have shown that the fraction of deuterons produced in the jet is 8--15$\%$, increasing with
increasing $p_{\rm T}$, while the majority of the deuterons are
produced in the underlying event. 

\end{document}